\documentclass[11pt]{article}
\usepackage{aaspp4}

\newcommand{\mjybm}{\mbox{mJy~beam${}^{-1}$}}

\newcommand{\aips}{{\textsc{aips}}}

\newcommand{\mion}[2]{\mathrm{#1}\:\mathrm{\expandafter\uppercase\expandafter{\romannumeral#2}\relax}}

\received{1997 June 5}
\accepted{1997 November 14}
\journalid{115}{1998 April}

\lefthead{Lazio \& Cordes}
\righthead{Angular Broadening Toward the Anticenter}
\begin{document}

\title{The Radial Extent and Warp of the Ionized Galactic Disk. I.\\
	A VLBA Survey of Extragalactic Sources \\
	Toward the Anticenter}

\author{T.~Joseph W.~Lazio\altaffilmark{1,2} \& James M.~Cordes}
\affil{Department of Astronomy and National Astronomy \& Ionosphere
	Center, Cornell University, Ithaca, NY  14853-6801; \\
	lazio@spacenet.tn.cornell.edu, cordes@spacenet.tn.cornell.edu}
\authoraddr{T. Joseph W. Lazio
            NRL, Code 7210
            4555 Overlook Ave. SW
            Washington, DC  20375-5351;
	    lazio@rsd.nrl.navy.mil}
\altaffiltext{1}{NRC/NRL Post-doctoral Associate}
\altaffiltext{2}{Current address: Remote Sensing Division, NRL,
	Code~7210, 4555 Overlook Ave. SW, Washington, DC  20375-5351;
	lazio@rsd.nrl.navy.mil}

\begin{abstract}
We report multifrequency Very Long Baseline Array observations of
twelve active galactic nuclei seen toward the Galactic anticenter.
All of the sources are at $|b| < 10\arcdeg$ and seven have $|b| <
0\fdg5$.  Our VLBA observations can detect an enhancement in the
angular broadening of these sources due to an extended \ion{H}{2}
disk, if the orientation of the \ion{H}{2} disk in the outer Galaxy is
similar to that of the \ion{H}{1} disk.  Such an extended \ion{H}{2}
disk is suggested by the \ion{C}{4} absorption in a quasar's spectrum,
the appearance of \ion{H}{1} disks of nearby spiral galaxies, and
models of Ly$\alpha$ cloud absorbers and the Galactic fountain.  We
detect eleven of the twelve sources at one or more frequencies; nine
of the sources are compact and suitable for an angular broadening
analysis.  A preliminary analysis of the observed angular diameters
suggests that the \ion{H}{2} disk does not display considerable
warping or flaring and does not extend to large Galactocentric
distances ($R \gtrsim 100$~kpc).  A companion paper (Lazio \& Cordes~1997)
combines these observations with those in the literature and presents
a more comprehensive analysis.
\end{abstract}


\section{Introduction}\label{sec:intro}

Refractive index fluctuations in the interstellar plasma, caused by
electron density fluctuations, cause compact radio sources to be
angularly broadened (\cite{r90}).  Interstellar angular broadening has
been observed for pulsars (e.g., Gwinn, Bartel, \& Cordes~1993),
masers (e.g., \cite{fdcv94}), and active galactic nuclei (e.g., Fey,
Spangler, \& Cordes~1991).  Though AGNs typically display intrinsic
structure comparable to the level of angular broadening observed on
many lines of sight ($\sim$ milliarcseconds), surveys of angular
broadening tend to focus on AGNs because they are strong sources,
uniformly distributed on the sky, and suffer from none of the distance
ambiguities normally associated with Galactic sources.  Nevertheless,
such surveys (e.g., \cite{dtbbbc84}; Fey, Spangler, \& Mutel~1989;
\cite{fsc91}) have been biased toward the inner Galaxy because of the
general enhancement of interstellar scattering seen toward the inner
Galaxy (\cite{dtbbbc84}; Cordes, Weisberg, \& Boriakoff~1985;
\cite{cwfsr91}; Taylor \& Cordes~1993, hereinafter \cite{tc93}).

At Galactic latitudes $|b| > 10\arcdeg$, the angular diameter at 1~GHz
due to interstellar scattering of a point source is approximately
$1\,\mathrm{mas}\,(\sin |b|)^{-1/2}$ (\cite{d-sr76}).  At low
latitudes the scattering becomes stochastic and its magnitude depends
upon whether the line of sight intersects a clump or localized region
of intense scattering (\cite{dtbbbc84}; Cordes et al.~1985;
\cite{cwfsr91}; \cite{tc93}).  At extremely low latitudes $|b| <
1\arcdeg$, sources toward the anticenter may show enhanced broadening
due to an extended \ion{H}{2} disk of the Galaxy.

The \ion{H}{1} disk of the Galaxy extends to a Galactocentric distance
of 25--30~kpc (\cite{b92}).  There is a small body of evidence that
hints that the radial extent of the \ion{H}{2} disk may equal or
exceed that of the \ion{H}{1} disk:
\begin{itemize}
\item Savage, Sembach, \& Lu~(1995) find \ion{C}{4} absorption along
the line of sight to H~1821$+$643 ($\ell = 94\arcdeg, b=27\arcdeg$).
Among the velocity components contributing to this absorption is
low-density ($n \sim 5.6 \times 10^{-3}$~cm${}^{-3}$), warm ($T \sim
10^4$~K) gas at a velocity of $-120$~km~s${}^{-1}$, corresponding to a
kinematic Galactocentric distance of 25~kpc.

\item The \ion{H}{1} disks of nearby galaxy are truncated at radii of
order 25--50~kpc, at which the surface density drops to
$N_{\mion{H}{1}} \lesssim 2 \times 10^{19}$~cm${}^{-2}$ (Corbelli,
Schneider, \& Salpeter~1989; \cite{v91}).  This truncation is observed
even for galaxies without nearby companions and likely occurs where
the disks become optically thin to the intergalactic ionizing flux
(\cite{ecees93}).  Charlton, Salpeter, \& Hogan~(1992) have proposed
that at least some of the low-redshift Ly$\alpha$ clouds seen in
quasar spectra may be due to residual \ion{H}{1} in extended, nearly
fully ionized disks of normal spiral galaxies.  Our Galaxy would then
be a prototypical, $z = 0$ absorber.

\item Material blown out of the Galactic disk by the action of
clustered supernovae may account for a fraction of high-velocity
clouds and later return to the disk, forming a Galactic fountain
(\cite{sf76}; \cite{b80}; \cite{hb90}; \cite{s90}; \cite{k91}).
Models of high-velocity clouds often require the material to be
supported by gas pressure at large Galactocentric radii, $R \gtrsim
25$~kpc (e.g., \cite{b80}).
\end{itemize}
The line of nodes of the \ion{H}{1} disk is approximately constant
with Galactocentric radius and centered near a Galactic longitude of
170\arcdeg.  Presuming that the \ion{H}{2} disk is oriented similarly
to the \ion{H}{1} disk, low-latitude sources toward the anticenter
would sample the longest path lengths through the \ion{H}{2} disk and
would be expected to show the largest enhancement from this extended
disk.

Figure~\ref{fig:angle} illustrates the enhanced angular broadening
expected as a function of the $e^{-1}$ radial scale length $A_1$.  A
recent effort to constrain $A_1$ (\cite{tc93}) found comparable fits
for $A_1 \approx 20$--50~kpc.  This large range is allowed because of
the dearth of anticenter scattering measurements.

In this paper we report multifrequency Very Long Baseline Array
observations of twelve anticenter sources, seven of which have $|b| <
0\fdg5$.  As Fig.~\ref{fig:angle} illustrates, the nominal resolutions
of the VLBA are such that 18~cm observations are sensitive to scale
lengths of $A_1 \gtrsim 100$~kpc and~90~cm observations should detect
scattering even if $A_1 \lesssim 10$~kpc.  In \S\ref{sec:ac.observe}
we describe the observations and data reduction, in
\S\ref{sec:sources} we present the results for individual sources, and
in \S\ref{sec:conclude} we present a preliminary analysis of the
angular diameters.  A companion paper (Lazio \& Cordes~1997,
hereinafter \cite{lc97}) combines these observations with the other
scattering measurements in the literature and uses a likelihood
analysis to constrain the distribution of scattering the outer Galaxy.

\section{Observational Program}\label{sec:ac.observe}

The sources observed in this project were selected from the Northern
Sky Catalogs (20~cm: \cite{wb92}, catalog acronym WB; 6~cm: Becker,
White, \& Edwards~1991, catalog acronym BWE) and the 6~cm MIT-Green
Bank survey (\cite{blbhm86}; \cite{lhclcb90}; \cite{glhclb90}).  Our
selection criteria were that the source have a flat spectrum with a
spectral index $|\alpha| \le 0.5$, that it be classified as point-like
to the Green Bank 100~m telescope, and that it be unlikely to be a
Galactic radio source (\cite{tg83}; \cite{gt86}).  Sources were also
selected to lie within a cross-shaped region in the Galactic
anticenter described by $150\arcdeg < \ell < 210\arcdeg$ and $|b| <
0\fdg5$ or $\ell \approx 180\arcdeg$ and $|b| < 10\arcdeg$.  Sources
near or behind known \ion{H}{2} regions (\cite{s59}) or supernova
remnants (\cite{g86}; \cite{g91}), objects which might enhance the
scattering, were excluded.

Because the above surveys were performed with single-dish telescopes,
the source positions were not sufficiently accurate for \hbox{VLBI}.  We
therefore undertook a VLA survey to refine source positions.  This
survey also enabled us to exclude those sources which are not compact
on arcsecond scales.

\subsection{VLA survey}\label{sec:vlasurvey}

In the surveys listed above, we identified 20 sources potentially
suitable for a VLBI angular broadening study; these are tabulated in
Table~\ref{tab:vlasources}.  On 1993 April~23, we observed these
sources with the VLA in the B configuration.  A 12.5~MHz bandpass was
used with the IFs centered on 1366 and 1446~MHz.  Snapshot
observations of duration 7~min.\ were obtained.  Observations of 3C286
and 3C147 were used to set the flux density scale and frequent
observations of 0552$+$398 were used to calibrate the visibility
phases.

The data were edited, calibrated, and imaged in the standard fashion
within \aips.  Images of those sources judged unsuitable for the VLBI
program are presented in Fig.~\ref{fig:vlasources}; we also show the
VLA image of the one source not detected in the VLBI program.  We
defer until \S\ref{sec:sources} the images of the sources observed in
the VLBI component of this program.

The fluxes given in Table~\ref{tab:vlasources} are derived from our
VLA observations.  Compact sources were fit with a single gaussian.
Doubles could typically be fit with two gaussians, in which case the
coordinates and brightness are for the stronger component while the
flux is the sum of the flux from both components.  For sources which
could not be fit by gaussians, the flux was derived by summing those
image pixels for which the brightness exceeded 2$\sigma$.  One source
we observed, 87GB~0600$+$3011, has no flux data listed in
Table~\ref{tab:vlasources}.  We could identify no source brighter than
10~\mjybm\ at this location.  This source appears in both the 6~cm BWE
and~20~cm WB catalogs with a flux $S \approx 300$~mJy.  We conclude
that it is likely to be largely resolved out to the VLA and is
probably a Galactic \ion{H}{2} region.  It is unlikely to be variable
because the observations from which the WB and BWE catalogs were
assembled were separated by 4~yr, 1983 to 1987.

\subsection{VLBA observations}\label{sec:vlbaobs}

Twelve sources were selected for further VLBI observations from the
VLA survey.  Of these, seven have Galactic latitudes $|b| < 0\fdg5$;
Dennison et al.~(1984) had three sources with $|b| < 1\arcdeg$.  These
very low-latitude sources are important because the lines of sight to
these sources traverse 50~kpc or more before exceeding the extended
component's scale height in the inner Galaxy, 0.88~kpc.  The remaining
five sources are useful in assessing the presence of any flaring or
warping of the outer ionized disk.  As a control source we also
observed 0611$+$131, a source in Dennison et al.'s~(1984) survey.  The
log of VLBI observations is given in Table~\ref{tab:vlbaobs}.

\subsubsection{6 and 18~cm Observations}\label{sec:obs6_18}

The first set of observations was conducted at 6 and 18~cm.  The 6~cm
observations allow determination of the intrinsic milliarcsecond
structure of the sources, while the 18~cm observations might be
capable of just detecting scattering if the Galactic disk extends to
100~kpc or more, viz.\ Fig.~\ref{fig:angle}.

The observations were conducted on 1994 May~6 and~7 in two 11~h
sessions.  Observations at two hour angles, of duration 6~min., were
obtained for each source at both wavelengths.  Dual polarization was
recorded in two 8-MHz IFs for a total bandwidth of 16~MHz.  The array
was composed of all of the VLBA antennas and the phased \hbox{VLA}.

Editing of the data was performed using station-supplied logs;
additional editing, mostly near scan boundaries, was also performed
later.  Amplitude calibration for the VLBA antennas was performed
using station-supplied $T_{\mathrm{sys}}$ measurements; for baselines
including the VLA, a source flux density is required.  At 18~cm we
estimated this quantity from our VLA observations.  At 6~cm we used
the fluxes from the BWE catalog (\cite{bwe91}), except for 0611$+$131,
for which we estimated a flux of 0.23~Jy from its spectrum
(\cite{dtbbbc84}; \cite{gr92}).  Fringe fitting was performed in two
steps.  First the delays across individual IFs due to the electronics
were determined using a short section of a scan on 0552$+$398.  These
delays were applied to all sources, then the rates, delay across the
IFs, and residual delay within the IFs were determined.  Corrections
for the shape of the bandpass provided little improvement in the
visibility phases or amplitudes and were not applied.

\subsubsection{90~cm}\label{sec:90}

The second set of observations was conducted at 90~cm.  Scattering
diameters should be detectable at this wavelength unless the scale
length of the Galaxy's ionized disk is less than 10~kpc in size,
viz.~Fig.~\ref{fig:angle}.

The observations were conducted in a single 12~h session on 1995
November~27 using the full VLBA and the phased \hbox{VLA}.  All of the
sources from the earlier VLBI session, except 87GB~0526$+$2458, were
observed.  The structure of this source at 6 and~18~cm, viz.\
Fig.~\ref{fig:J0529+2500}, is suggestive of a compact symmetric object
(\cite{cpruxm92}).  We therefore judged it to be unsuitable for
scattering measurements and excluded it from further observations.
Each source was observed over two hour angles for a total time on
source of 30~min.  The sources 3C147, 3C286, and 3C454.3 were also
observed to assist with fringe finding.

Only left-circular polarization was recorded.  The VLBA stations
recorded eight 4-MHz IFs, with 16 channels per IF, for a total
bandwidth of 32~MHz; the VLA hardware and radio frequency interference
(RFI) environment allowed only four IFs to be recorded.

Initial editing was performed using station-supplied logs.  At 90~cm,
RFI is problematic; extensive additional editing was performed to
excise \hbox{RFI}.  Amplitude calibration and fringe finding followed a
similar procedure as the 6 and~18~cm observations.  The single-IF
delays were found by utilizing a 2~min.\ section from one scan of
3C147.  RFI made it difficult to isolate a section of a scan
containing the full frequency response of the antennas, thus no
correction for the shape of the bandpass was applied.

\section{Individual Sources}\label{sec:sources}

After the delay and rate solutions were applied, the data were
averaged across the full bandpass (16~MHz at~6 and~18~cm and 32~MHz
at~90~cm) and in time.  At~6 and~18~cm the averaging time was
typically 30~s and at~90~cm it was 10~s.  Longer averaging times were
used if a source was not detected within these initial averaging
times.  Of the twelve sources observed, we detect all but one,
WB~0616$+$1522 (viz.\ Fig~\ref{fig:vlasources}), at one or more
wavelengths.

Gain fluctuations remaining in the data were then removed via a few to
several iterations of self-calibration (e.g., \cite{w95}).  For the
weaker sources, only the gain phases were corrected; both gain
amplitude and phase corrections were determined for the stronger
sources.  In the initial self-calibration iteration we used a
point-source model to correct the phases only.  Thereafter we used the
images produced from the self-calibrated data as the input model.  We
ceased the self-calibration iterations when the off-source rms in an
image was within a factor of 2--3 of the thermal limit, which is
0.6~\mjybm\ at 6 and~18~cm and 2~\mjybm\ at 90~cm.  We did not attempt
to reach the noise level given our limited hour angle coverage.

A summary of the sources is given in Table~\ref{tab:sizes} and
Figs.~\ref{fig:spectra} and \ref{fig:sizes}.
Figure~\ref{fig:spectra} shows the spectrum of each source detected at
two or three of the VLBI observation wavelengths.  Because of our
limited hour angle coverage, we were unable to test our amplitude
calibration via such means as $u$-$v$ plane crossings.  The \textit{a
priori} amplitude calibration at the VLBA should be accurate to 10\%
(\cite{w96}).  Figure~\ref{fig:sizes} shows the angular diameter as a
function of wavelength for each source appearing compact at two or
three of the VLBI observing wavelengths.

Our primary aim for undertaking this survey was to find extragalactic
sources whose angular diameters show evidence of radio-wave scattering
and could therefore be used to constrain the size of the \ion{H}{2}
disk (\cite{lc97}).  Not all of the sources we have observed are suitable
for our intended analysis.  In this section we present the individual
sources and an assessment of their use for further analysis.  For
sources with complex structure, we show the image produced from the
self-calibrated visibility data.  For sources showing simple
structure, i.e., one or two gaussian components, we show the
visibility data.  If the source can be modelled with a single
gaussian, we superpose that model on the data.

Because of the limited hour angle coverage, the angular diameters in
Table~\ref{tab:sizes} are found from fits of circular gaussians to the
$u$-$v$ data.  For a subset of the sources, we also performed fits of
an elliptical gaussian.  We focussed on the 90~cm observations as
these observations have the longest duration and, therefore, the most
complete $u$-$v$ coverage.  The 90~cm observations are also the most
sensitive to scattering-induced anisotropy, though other causes such
as intrinsic structure or $u$-$v$ coverage can also produce
anisotropic image shapes.  We restricted fits of an elliptical
gaussian to those sources with $S > 0.5$~Jy at~90~cm, so that the data
would have a high signal-to-noise ratio.  Four sources---87GB~0512$+$2627,
87GB~0547$+$3044, 87GB~0600$+$2957, and 0611$+$131---meet these
criteria.  

We characterize the elliptical fits by the image elongation
parameter $e_{\mathrm{s}}$ (Romani, Narayan, \& Blandford~1986)
\begin{eqnarray}
e_{\mathrm{s}} & = & \frac{a-b}{\sqrt{ab}} \nonumber \\
 & = & \frac{1-b/a}{\sqrt{b/a}}
\label{eqn:elongate}
\end{eqnarray}
where $a$ and~$b$ are the major and minor axes of the image.  Romani
et al.~(1986) pointed out that even though the expected image shape
from an isotropic scattering medium is circular, a single realization
of the image shape can produce a non-zero $e_{\mathrm{s}}$.  They
provide expressions for the rms value of $e_{\mathrm{s}}$, $\sigma_e$,
in terms of the relevant observational and scattering parameters
(their Table~1).  For our 90~cm observational program and using the
\cite{tc93} model to describe scattering in the anticenter, we
estimate $\sigma_e \approx 0.03$ for anticenter sources.  Values of
$e_{\mathrm{s}}$ significantly larger than this indicate
(1)~scattering toward the anticenter is stronger than that described
by the \cite{tc93} model (i.e., larger $A_1$); (2)~the density
fluctuations, which are responsible for interstellar scattering, are
described by a so-called ``steep'' power-law spatial spectrum in the
anticenter (\cite{rnb86}; \cite{r90}); (3)~the scattering medium is
anisotropic; or (4)~effects not related to scattering, such as those
cited above, are important.  We discuss the fits individually, but we
shall conclude that $u$-$v$ coverage or intrinsic structure are more
likely than scattering to be responsible for the anisotropic images
shapes that we find.

\subsection{87GB~0433$+$4706}\label{sec:j0436+4712}

Although appearing compact to the VLA at 6 and~20~cm
(Fig.~\ref{fig:J0436+4712}; \cite{f86}), we detect this source only at
90cm in the VLBI observations.  The lack of a VLBI detection at 6
and~18~cm may be due to scheduling and correlator constraints which
resulted in this source not being observed by the VLA, the most
sensitive element in our VLBI array.  

Because its 90~cm structure can be fit by a single gaussian, we use
this source in our analysis of scattering.  The diameter of this
source is a factor of ten larger than two nearby sources, 87GB~0451$+$4309
(see below) and 0503$+$467 (\cite{smbc86}), even though these sources
are closer to the supernova remnant HB9.  This large diameter may be
an indication of anomalously strong scattering or of unresolved
structure (like 0611$+$131, \S\ref{sec:0611+131}).

\subsection{87GB~0451$+$4309}\label{sec:j0455+4313}

Like 87GB~0433$+$4706 this source appears compact to the VLA at 20~cm, but
is detected only at 90~cm, Fig.~\ref{fig:J0455+4313}.  The same cause
is likely responsible for this lack of detection as the VLA was unable
to observe this source.  

Because its 90~cm structure can be fit by a single gaussian, we will
use this source in our analysis.  The diameter of this source, 5~mas
scaled to 1~GHz assuming $\theta \propto \lambda^{2.2}$, compares
favorably with the 4~mas diameter Spangler et al.~(1986) find for
0503$+$467.

\subsection{87GB~0512$+$2627}\label{sec:j0516+2630}

We detect this source at 18 and~90~cm, Fig.~\ref{fig:J0516+2630}.  It
can be fit by a single gaussian at both wavelengths and we shall use
this source in our analysis.

This was one of the sources to which we also fit an elliptical
gaussian.  The image elongation parameter is $e_{\mathrm{s}} = 0.11$
(axial ratio~$b/a = 0.899$).  This value is significantly in excess of
the rms value for isotropic scattering.  The orientation of the
source's major axis is aligned approximately with the orientation of
the longest baselines in the $u$-$v$ plane.  We conclude that the
image shape is likely to be dominated by the incomplete $u$-$v$
coverage, and that a circular gaussian is an equally acceptable
description of the source shape.

\subsection{87GB~0526$+$2458}

At 20~cm, this source appears compact to the VLA,
Fig.~\ref{fig:J0529+2500}.  At 18~cm our VLBI observations resolve the
source into a compact symmetric object (\cite{cpruxm92}) having two
components separated by approximately 60~mas.  The weaker of these is
not visible in our 6~cm image.  Because of its structure, we judged it
to be unsuitable for an angular broadening analysis and did not
observe it at 90~cm.

\subsection{87GB~0537$+$3059}

To the VLA, this source appears compact, Fig.~\ref{fig:J0541+3046}.
At 90~cm the source shows a simple structure.  At 18~cm two components
are visible, a compact core-like component and a larger halo.  At 6~cm
the source is detected on only the shortest baseline, PT-\hbox{VLA}.  From
the two shortest baselines, PT-VLA and LA-VLA, we constrain its 6~cm
diameter to $57\,\mathrm{mas} < \theta < 284\,\mathrm{mas}$.  We shall use
the diameters derived from the 90~cm data and the 18~cm core component
in our analysis.

\subsection{87GB~0547$+$3044}

At 20~cm, the source has a compact core with a jet extending
approximately 30\arcsec\ to the west, Fig.~\ref{fig:J0550+3045}.  In
our VLBI observations we detect the core component at all three
wavelengths.  However, at 6~cm, it is detected on only the two
shortest baselines, PT-VLA and LA-VLA, so that we can constrain its
diameter to be only $\theta < 57$~mas.  We use only the diameters
derived from the 18 and~90~cm data in our scattering analysis.

This was one of the sources to which we also fit an elliptical
gaussian.  The image elongation is $e_{\mathrm{s}} = 0.17$ ($b/a =
0.842$).  Like for 87GB~0512$+$2627, \S\ref{sec:j0516+2630}, however,
the image shape is likely to be dominated by the incomplete $u$-$v$
coverage as the orientation of the source's major axis is aligned
approximately with the orientation of the longest baselines in the
$u$-$v$ plane.

\subsection{87GB~0558$+$2325}

We detect this low-latitude source at all three wavelengths,
Fig.~\ref{fig:J0601+2324}.  It is compact and we use it in our
analysis.

\subsection{87GB~0600$+$2957}\label{sec:J0603+2957}

This source is detected at all three wavelengths,
Fig.~\ref{fig:J0603+2957}.  At 6 and~18~cm, two components can be
identified---a compact one with a flux density of 150--250~mJy and a
secondary component approximately 2~mas to the west with a flux of
100--150~mJy.  At 90~cm, only one component is present.  For our
scattering analysis, we shall use only the more compact component.

This was one of the sources to which we also fit an elliptical
gaussian.  The image elongation is $e_{\mathrm{s}} = 0.44$ ($b/a =
0.649$).  This elongation is likely to reflect intrinsic structure.
First, the wavelength dependence of the angular diameter is $\theta
\propto \lambda^{0.3}$, as opposed to the $\lambda^2$ dependence
expected for angular broadening.  Second, an image of the source
at~18~cm shows a comparable elongation, though with a misalignment of
the position angles.  The 18~cm image has a position angle of
approximately $-10\arcdeg$ while the fit to the $u$-$v$ data gives a
position angle of approximately 30\arcdeg.

In \cite{lc97} we show that the scattering diameter for this source is
small enough to alter the estimates of the scale height of the ionized
disk in the outer Galaxy by 50\%.  Possible explanations for a small
scattering diameter are that the distribution of scattering material
is patchy or that the source is Galactic.  However, we have found no
characteristic of the source which would cause us to favor a Galactic
classification over extragalactic: Its brightness temperature is about
$5 \times 10^9$~K; its morphology is consistent with that of
extragalactic sources, a central core with a secondary component;
there are no X-ray sources within 1\arcdeg; and the nearest pulsar is
PSR~B0609$+$37.  We discuss these possibilities in more detail in
\cite{lc97}.

Condon et al.(1983) were unable to find an optical counterpart for
this source, though the field does show signs of obscuration.

\subsection{87GB~0621$+$1219}

This source is detected at all three wavelengths and can be modelled
with a single gaussian component, Fig.~\ref{fig:J0623+1218}.  We shall
use this source in our analysis.

\subsection{87GB~0621$+$3206}

This source is detected at 18 and~90~cm, Fig.~\ref{fig:J0624+3205}.
At 18~cm the source is resolved into two components, with a core-jet
morphology.  The visibility data show significant deviations from a
gaussian, particularly at short baselines, and there is possibly a
extended, nearly resolved-out component approximately 400~mas from the
core.  At 90~cm the source has two components, separated by 400~mas.
Because of its complex structure, we shall not use this source in our analysis.

\subsection{87GB~0622$+$1153}

This source is detected at 18 and~90~cm, Fig.~\ref{fig:J0625+1150}.
At 90~cm the source can be modelled with a single component.  At
18~cm, we detect the source only on the PT-LA-VLA triangle.  The
visibility amplitude on the PT-VLA baseline, the shortest baseline, is
lower than that on the LA-VLA and PT-LA baselines, possibly indicating
that the source structure at 18~cm may be more complex than at 90~cm.
Given the paucity of data at 18~cm, however, we shall use only the
90~cm diameter in our scattering analysis.

\subsection{0611$+$131}\label{sec:0611+131}

This source was observed by Dennison et al.~(1984) and was included in
our observations as a control source.  Our spectrum is in good
agreement with that obtained by Dennison et al.~(1984), though the
wavelengths from which they derive their spectrum differ somewhat from
ours.  Also, we find a flux 10--15\% lower than Dennison et
al.~(1984).  Our 90~cm source diameter is 41.4~mas, in good agreement
with their upper limit of 40~mas at 75~cm.  However, as our 6
and~18~cm images show, Fig.~\ref{fig:0611+131}, these low-frequency
source diameters are unlikely to be that of a single component.  At
higher frequencies, a second component is seen to the north,
approximately 35~mas away from the compact component we identify as
the core.  Extended emission is also seen either partially or entirely
linking these two components.  We shall therefore exclude this source
 from our analysis (\cite{lc97}).

This was one of the sources to which we also fit an elliptical
gaussian.  The image elongation is $e_{\mathrm{s}} = 0.66$ ($b/a =
0.521$).  This is the most highly anisotropic of the four sources for
which we fit elliptical gaussians.  The anisotropy of this source is
likely to be dominated by the intrinsic structure.  Our higher
resolution observations at~6 and~18~cm reveal an elongated source
structure with a comparable elongation.  There is some misalignment
between the image shapes, however.  The 6 and~18~cm images show the
source to be elongated in the north-south direction, a position angle
of 0\arcdeg; the fit to the $u$-$v$ data at~90~cm finds a position
angle of 24\arcdeg.

\section{Preliminary Constraints on the Ionized Disk}\label{sec:conclude}

The measured angular diameters are summarized in
Table~\ref{tab:sizes}.  They are in the range 50--600~mas at 90~cm,
1--150~mas at 18~cm, and 0.4--5~mas at 6~cm.  The latitude range of
the sources is $|b| < 10\arcdeg$.

At high latitudes the expected scattering diameter of extragalactic
sources is $12\,\mathrm{mas}\,\lambda^2(\sin |b|)^{-1/2}$ for
$\lambda$ in meters (\cite{d-sr76}).  Of the three wavelengths at
which we observed, the 90~cm angular diameters will contain the
largest contribution from scattering.  Scaling the high latitude
expression to 327~MHz, the expected scattering diameters are
approximately 20--75~mas for sources with $|b| = 1\arcdeg$ to
10\arcdeg.

Of the ten sources for which we have measured an angular diameter at
90~cm, seven of them have diameters smaller than 90~mas, viz.\
Table~\ref{tab:sizes}, comparable to that expected if the Galaxy's
scattering material is confined to a flat disk.  Of the remaining
three sources, we cannot assess the amount to which intrinsic
structure contributes to the observed diameters of 87GB~0433$+$4706
and 87GB~0622$+$1153 because they were detected at only one frequency,
and scattering probably does not influence the size of
87GB~0537$+$3059h because it is the halo of the compact component.
The close correspondence between the observed angular diameters and
that extrapolated from high latitudes suggests that the ionized disk
is not strongly warped like \ion{H}{1} disk.  It may also indicate
that the scattering material does not extend to large Galactocentric
distances ($R \gtrsim 100$~kpc).  We defer a more comprehensive
analysis to \cite{lc97} where we combine these observations with those
in the literature and use a likelihood analysis to constrain the
distribution of scattering the outer Galaxy.

\acknowledgements
This research has made use of the Simbad database, operated at the
CDS, Strasbourg, France.  The Very Large Array (VLA) and the Very Long
Baseline Array (VLBA) are facilities of the National Science
Foundation operated by the National Radio Astronomy Observatory
under cooperative agreement by Associated Universities, Inc.  Many
people at the NRAO were instrumental in assisting with these
observations.  We thank A.~Beasley for assistance with the L- and
C-band observations; C.~Janes, R.~Simon, and J.~Wrobel spent
considerable time determining the RFI environment, particularly for
the VLA, and assisting with the production of SCHED files.  We thank
B.~Clark for scheduling us on a Thanksgiving weekend for the 90~cm
observations.  This research was supported by NASA GRO grants
NAG~5-2436 and NAG~5-3515 and NSF grant AST-9528394.

\clearpage

\begin{figure}
\caption[Angular Broadening in the Anticenter from the Taylor-Cordes Model]
{The angular broadening for an extragalactic source observed toward
the anticenter as a function of $A_1$, the Galactocentric scale length
in the Taylor-Cordes model.  The curves are labelled by the
observation frequency in GHz.  The VLBA's nominal resolutions at these
two frequencies are shown as dotted lines.  For comparison, we also
show the nominal resolution for the space VLBI satellite \hbox{HALCA}.}
\label{fig:angle}
\end{figure}

\begin{figure}
\caption[VLA Images of Sources Not Observed in the VLBI Observations]
{VLA images of sources not observed or not detected in the VLBI
observations.  Contour levels are in units of 0.6~\mjybm, the thermal
noise level, and are $-2$, 3, 5, 10, 20, \dots; the size of the beam
is indicated in the lower left of the plot.
(\textit{a}) 87GB~0504$+$2402; 
(\textit{b}) 87GB~0512$+$2455; 
(\textit{c}) 87GB~0514$+$3730;
(\textit{d}) 87GB~0559$+$2911;
(\textit{e}) WB~0602$+$2102;
(\textit{f}) WB~0616$+$1522, this source was observed, but not detected
in the VLBI observations;
(\textit{g}) WB~0623$+$3227; and
(\textit{h}) 87GB~0631$+$0907.}
\label{fig:vlasources}
\end{figure}

\begin{figure}
\caption[VLBI Source Spectra]
{The spectra of the VLBI sources detected at multiple frequencies.}
\label{fig:spectra}
\end{figure}

\begin{figure}
\caption[VLBI Source Diameters]
{The \emph{apparent} source diameters for those VLBI sources which could
be fit with a single gaussian.  The top panel shows the expected
slope for a source in which scattering dominates at all frequencies.}
\label{fig:sizes}
\end{figure}

\begin{figure}
\caption[87GB~0433$+$4706]
{87GB~0433$+$4706.  (\textit{a}) The 20~cm VLA image.  Levels are
0.6~\mjybm\ $\times$ $-2$, 3, 5, 10, 20, \ldots.  In this VLA image and
subsequent VLA and VLBI images, the contours are given in units of the
off-source rms level.  (\textit{b}) The 90~cm visibility data for
87GB~0433$+$4706.  In order to avoid a noise bias, we display the real
part of the visibility; a 617~mas gaussian model is superposed.}
\label{fig:J0436+4712}
\end{figure}

\begin{figure}
\caption[87GB~0451$+$4309]
{87GB~0451$+$4309. (\textit{a}) The 20~cm VLA image.  The contour
levels are 0.6~\mjybm\ $\times$ $-2$, 3, 5, 10, 20, \ldots.  The beam is
shown in the lower left.  (\textit{b}) The 90~cm (real part of the)
visibility data of 87GB~0451$+$4309 showing a 65~mas gaussian model
superposed.}
\label{fig:J0455+4313}
\end{figure}

\begin{figure}
\caption[87GB~0512$+$2627]
{87GB~0512$+$2627. (\textit{a}) The 20~cm VLA image.  Contour levels
are 0.6~\mjybm\ $\times$~$-2$, 3, 5, 10, 20, \ldots.  The beam is
shown in the lower left.
(\textit{b}) The 90~cm visibility data of 87GB~0512$+$2627 with an
87~mas gaussian model superposed.
(\textit{c}) The 18~cm visibility data of 87GB~0512$+$2627 with a
35~mas gaussian model superposed.}
\label{fig:J0516+2630}
\end{figure}

\begin{figure}
\caption[87GB~0526$+$2458]
{87GB~0526$+$2458, because of its compact symmetric morphology at the
higher frequencies, this source was not observed at 90~cm.
(\textit{a}) The 20~cm VLA image.  Contour levels are 0.6~\mjybm\
$\times$~$-2$, 3, 5, 10, 20,~\ldots.  The beam is shown in the lower
left.
(\textit{b}) 18~cm VLBI image of 87GB~0526$+$2458.  Contours are
1.6~\mjybm\ $\times$~$-2$, 3, 5, 10, 20,~\ldots.  The beam is shown in
the lower left.
(\textit{c})  6~cm VLBI image of 87GB~0526$+$2458.  Contours are
3.6~\mjybm\ $\times$~$-2$, 3, 5, 10, 20, \ldots.  The beam is shown in
the lower left.}
\label{fig:J0529+2500}
\end{figure}

\begin{figure}
\caption[87GB~0537$+$3059]
{87GB~0537$+$3059, at 6~cm, the source is detected on only the
shortest baseline, PT-VLA, and we do not show this baseline.
(\textit{a}) The 20~cm VLA image.  Contour levels are 0.6~\mjybm\
$\times$~$-2$, 3, 5, 10, 20,~\ldots.  The beam is shown in the lower
left.
(\textit{b}) The 90~cm visibility data of 87GB~0537$+$3059 with a
485~mas gaussian model superposed.
(\textit{c}) The 18~cm (real part of the) visibility data of
87GB~0537$+$3059.  The two dotted lines show 1.2 and~77~mas gaussian
components.  The solid line shows their sum.}
\label{fig:J0541+3046}
\end{figure}

\begin{figure}
\caption[87GB~0547$+$3044]
{87GB~0547$+$3044, at 6~cm, the source is detected on only the two
shortest baselines, PT-VLA and LA-VLA, which we do not show.
(\textit{a}) The 20~cm VLA image.  Contour levels are 0.6~\mjybm\
$\times$~$-2$, 3, 5, 10, 20,~\ldots.  The beam is shown in the lower
left.
(\textit{b}) The 90~cm visibility data of 87GB~0547$+$3044 with a
78~mas gaussian model superposed.
(\textit{c})  The 18~cm (real part of the) visibility data of
87GB~0547$+$3044 with a 48~mas gaussian model superposed.}
\label{fig:J0550+3045}
\end{figure}

\begin{figure}
\caption[87GB~0558$+$2325]
{87GB~0558$+$2325.  (\textit{a}) The 20~cm VLA image. Contours are
0.2~\mjybm\ $\times$ $-2$, 2, 3, 5, 10, \ldots.  The beam is shown in
the lower left.
(\textit{b})  The 90~cm visibility data of 87GB~0558$+$2325 with a
49~mas gaussian model superposed.
(\textit{c})  The 18~cm visibility data of 87GB~0558$+$2325 with a
2.84~mas gaussian model superposed.
(\textit{d})  The 6~cm visibility data of 87GB~0558$+$2325 with a
0.39~mas gaussian model superposed.}
\label{fig:J0601+2324}
\end{figure}

\begin{figure}
\caption[87GB~0600$+$2957]
{87GB~0600$+$2957.  (\textit{a}) The 20~cm VLA image.  Contour levels
are 0.6~\mjybm\ $\times$~$-2$, 2, 3, 5, 10, 20, \ldots.  The beam is
shown in the lower left corner.
(\textit{b})  The 90~cm visibility data of 87GB~0600$+$2957 with a
21~mas gaussian model superposed.
(\textit{c})  The 18~cm visibility data of 87GB~0600$+$2957, with a
1.5~mas gaussian model superposed; the additional flux at short
baselines is from a secondary component 2~mas to the west of the
central component.
(\textit{d})  The 6~cm visibility data of 87GB~0600$+$2957, with a
0.7~mas gaussian model superposed, the additional flux at short
baselines is from a secondary component 2~mas to the west of the
central component.}
\label{fig:J0603+2957}
\end{figure}

\begin{figure}
\caption[87GB~0621$+$1219]
{87GB~0621$+$1219.  (\textit{a}) The 20~cm VLA image.    Contour
levels are 0.6~\mjybm\ $\times$~$-2$, 2, 3, 5, 10, 20, \ldots.  The
beam is shown in the lower left corner.
(\textit{b})  The 90~cm (real part of the) visibility data of
87GB~0621$+$1219 with a 38~mas gaussian model superposed.
(\textit{c})  The 18~cm (real part of the) visibility data of
87GB~0621$+$1219 with a 2.8~mas gaussian model superposed.
(\textit{d})  The 6~cm (real part of the) visibility data of
87GB~0621$+$1219 showing a 1.1~mas gaussian model superposed.}
\label{fig:J0623+1218}
\end{figure}

\begin{figure}
\caption[]{87GB~0621$+$3206.  (\textit{a}) The 20~cm VLA image.
Contour levels are 0.6~\mjybm\ $\times$~$-2$, 2, 3, 5, 10, 20, \ldots.
The beam is shown in the lower left corner.
(\textit{b})  The 90~cm VLBI image of 87GB~0621$+$3206.  Contours are
10~\mjybm\ $\times$~$-2$, 3, 5, 10.
(\textit{c})  The 18~cm VLBI image of 87GB~0621$+$3206.}
\label{fig:J0624+3205}
\end{figure}

\begin{figure}
\caption[87GB~0622$+$1153]
{87GB~0622$+$1153, at 18~cm, the source is detected on only the
LA-PT-VLA triangle, which we do not show.
(\textit{a}) The 20~cm VLA image.  Contours are 0.6~\mjybm\
$\times$~$-2$, 2, 3, 5, 10, 20, \ldots.  The beam is shown in the
lower left.
(\textit{b})  The 90~cm visibility data of 87GB~0622$+$1153 with a
188~mas gaussian model superposed.}
\label{fig:J0625+1150}
\end{figure}

\begin{figure}
\caption[0611$+$131]
{0611$+$131.  (\textit{a}) The 20~cm VLA image.  Contours are 0.6~\mjybm
$\times$ $-2$, 2, 3, 5, 10, 20, \ldots.  The beam is shown in the lower left
corner.  (\textit{b})  The 90~cm visibility data of 0611$+$131 with a
41.4~mas gaussian model superposed.  (\textit{c})  The 18~cm VLBI image of 0611$+$131.  Contours are 1.1~\mjybm\
$\times$ $-2$, 3, 5, 10, 20, \ldots.  The beam is shown in the lower
left.  (\textit{d})  The 6~cm VLBI image of 0611$+$131.  Contour levels are 1.6~\mjybm\
$\times$ $-2$, 3, 5, 10, 20, \ldots.  The beam is shown in the lower left.}
\label{fig:0611+131}
\end{figure}

\clearpage

\begin{deluxetable}{lccccccc}
\tablecaption{VLA 20~cm Survey\label{tab:vlasources}}
\tablehead{\colhead{Name} 
	& \colhead{RA} & \colhead{Dec}
	& \colhead{$\ell$} & \colhead{$b$}
	& \colhead{$I$} & \colhead{$S$} \nl
        & \multicolumn{2}{c}{(J2000)}
	& \colhead{(\arcdeg)} & \colhead{(\arcdeg)}
	& \colhead{(mJy/beam)} & \colhead{(mJy)} }

\startdata
\multicolumn{7}{c}{$150\arcdeg < \ell < 210\arcdeg$, $|b| < 0\fdg5$} \nl

87GB~0433$+$4706 & 04 36 43.267 & 47 12 36.97 & 157.3 & $-$0.0 & \phn167 & \phn177 \nl
87GB~0451$+$4309 & 04 55 32.873 & 43 13 53.96 & 162.5 & $-$0.1 & \phn115 & \phn118 \nl
87GB~0514$+$3730 & 05 18 08.891 & 37 32 16.10 & 169.6 & $-$0.0 & \phn\phn31 & \phn\phn39 \nl
87GB~0537$+$3059 & 05 41 30.806 & 30 46 14.77 & 177.9 & \phs0.2 & \phn\phn94 & \phn162 \nl
87GB~0558$+$2325 & 06 01 47.362 & 23 24 53.40 & 186.5 & \phs0.3 & \phn206 & \phn210 \nl

\nl 

WB~0602$+$2102 & 06 05 23.357 & 21 03 42.96 & 189.0 & $-$0.1 & \phn\phn28 & \phn\phn78 \nl
WB~0616$+$1522 & 06 19 09.473 & 15 21 02.42 & 195.6 & \phs0.0 & \phn163 & \phn357 \nl
87GB~0621$+$1219 & 06 23 52.825 & 12 18 29.49 & 198.8 & $-$0.4 & \phn124 & \phn124 \nl
87GB~0622$+$1153 & 06 25 15.482 & 11 50 10.19 & 199.4 & $-$0.3 & \phn\phn64 & \phn\phn67 \nl
87GB~0631$+$0907 & 06 34 31.429 & 09 04 37.06 & 202.9 & \phs0.4 & \phn\phn\phn8 & \phn\phn28 \nl

\tableline
\tablebreak

\multicolumn{7}{c}{$\ell \approx 180\arcdeg$, $|b| < 10\arcdeg$} \nl

87GB~0504$+$2402 & 05 07 05.690 & 24 05 51.53 & 179.2 & $-$9.8 & \phn136 & \phn342 \nl
87GB~0512$+$2455 & 05 15 51.031 & 24 59 09.73 & 179.6 & $-$7.7 & \phn\phn33 & \phn605 \nl
87GB~0512$+$2627 & 05 16 44.279 & 26 30 46.99 & 178.5 & $-$6.6 & \phn174 & \phn175 \nl
87GB~0526$+$2458 & 05 29 10.149 & 25 00 51.53 & 181.3 & $-$5.1 & 1166 & 1166 \nl
87GB~0547$+$3044 & 05 50 14.635 & 30 45 46.40 & 178.9 & \phs1.8 & \phn305 & \phn330 \nl
						 
\nl						 
						 
87GB~0559$+$2911 & 06 02 39.558 & 29 11 05.00 & 181.6 & \phs3.3 & \phn\phn28 & \phn401 \nl
87GB~0600$+$2957 & 06 03 55.856 & 29 57 05.35 & 181.1 & \phs3.9 & \phn508 & \phn518 \nl
WB~0600$+$3011   & 06 03 46.596 & 30 11 02.50 & 180.9 & \phs4.0 & \nodata & \nodata \nl
87GB~0621$+$3206 & 06 24 57.308 & 32 05 04.94 & 181.3 & \phs8.9 & 1328 & 1364 \nl
WB~0623$+$3227   & 06 27 54.773 & 32 25 59.38 & 180.9 & \phs8.8 & \phn\phn51 & \phn108 \nl
\enddata
\end{deluxetable}

\clearpage

\begin{deluxetable}{lccccc}
\small
\tablecaption{VLBI Observing Log\label{tab:vlbaobs}}
\tablehead{ & & & & & \colhead{Observing} \\
	\colhead{Wavelength} & \colhead{Date} & \colhead{Array} &
	\colhead{Polarization} & \colhead{Bandwidth} & \colhead{Time}}

\startdata
\phn6~cm & 1994 May 5--6 & VLBA, phased VLA & RR \& LL & 16~MHz & 11${}^{\mathrm{h}}$ \nl

18~cm    & 1994 May 5--6 & VLBA, phased VLA & RR \& LL & 16~MHz & 11${}^{\mathrm{h}}$ \nl

90~cm    & 1995 Nov 27\phn & VLBA, phased VLA & LL & 32~MHz & 12${}^{\mathrm{h}}$ \nl

\enddata
\end{deluxetable}

\begin{deluxetable}{lcccccc}
\tablecaption{VLBI Source Diameters\tablenotemark{a}\label{tab:sizes}}
\tablehead{
	& \multicolumn{2}{c}{90~cm} 
	& \multicolumn{2}{c}{18~cm} 
	& \multicolumn{2}{c}{6~cm} \\
	\cline{2-3} \cline{4-5} \cline{6-7}
	\colhead{Name}
	& \colhead{Flux} & \colhead{$\theta$}
	& \colhead{Flux} & \colhead{$\theta$}
	& \colhead{Flux} & \colhead{$\theta$} \\
	& \colhead{(mJy)} & \colhead{(mas)}
	& \colhead{(mJy)} & \colhead{(mas)}
	& \colhead{(mJy)} & \colhead{(mas)}}

\startdata

87GB~0433$+$4706 & 323\phd\phn & 617\phd\phn    & \nodata     & \nodata            & \nodata           & \nodata \nl
87GB~0451$+$4309 & 234.7       & \phn65.4       & \nodata     & \nodata            & \nodata           & \nodata \nl
87GB~0512$+$2627 & 559.2       & \phn87\phd\phn & 116\phd\phn & \phm{$<$}\phn35\phd\phn\phn & \nodata           & \nodata \nl
87GB~0537$+$3059c & \nodata     & \nodata        & \phn22.6    & $<$\phn\phn1.2\phn & \phn\phn5\phd\phn & $<$264\phd\phn \nl
87GB~0537$+$3059h & 736\phd\phn & 485\phd\phn    & 165\phd\phn & \phm{$<$}\phn76.7\phn & \nodata           & \nodata \nl

\nl 

87GB~0547$+$3044 & 932\phd\phn & \phn78.2       & 172\phd\phn & \phm{$<$}\phn48\phd\phn\phn & \phn\phn5\phd\phn & $<$\phn57 \nl
87GB~0558$+$2325 &  \phn76.2   & \phn49\phd\phn & 212.2       & \phm{$<$}\phn\phn2.84       & 390.4             & \phm{$<$}\phn\phn0.39 \nl
87GB~0600$+$2957c & 540\phd\phn & \phn20.8       & 262\phd\phn & \phm{$<$}\phn\phn1.45       & 150.0             & \phm{$<$}\phn\phn0.73 \nl
87GB~0600$+$2957h & \nodata     & \nodata        & 158\phd\phn & \phm{$<$}\phn\phn6.50       & \phn97.3          & \phm{$<$}\phn\phn5.3\phn \nl
87GB~0621$+$1219 & 182.2       & \phn38.2       & \phn83.6    & \phm{$<$}\phn\phn2.8\phn    & \phn51.3          & \phm{$<$}\phn\phn1.14 \nl

\nl

87GB~0622$+$1153 & 119\phd\phn & 188\phd\phn    & \nodata     & \nodata            & \nodata & \nodata \nl

\nl

0611$+$131  & 643\phd\phn & \phn41.4        & \nodata & \nodata & \nodata & \nodata \nl

\enddata

\tablenotetext{a}{Diameters are tabulated only for those sources and
at those frequencies at which one or two gaussian components could be
fit to the data.}
\end{deluxetable}

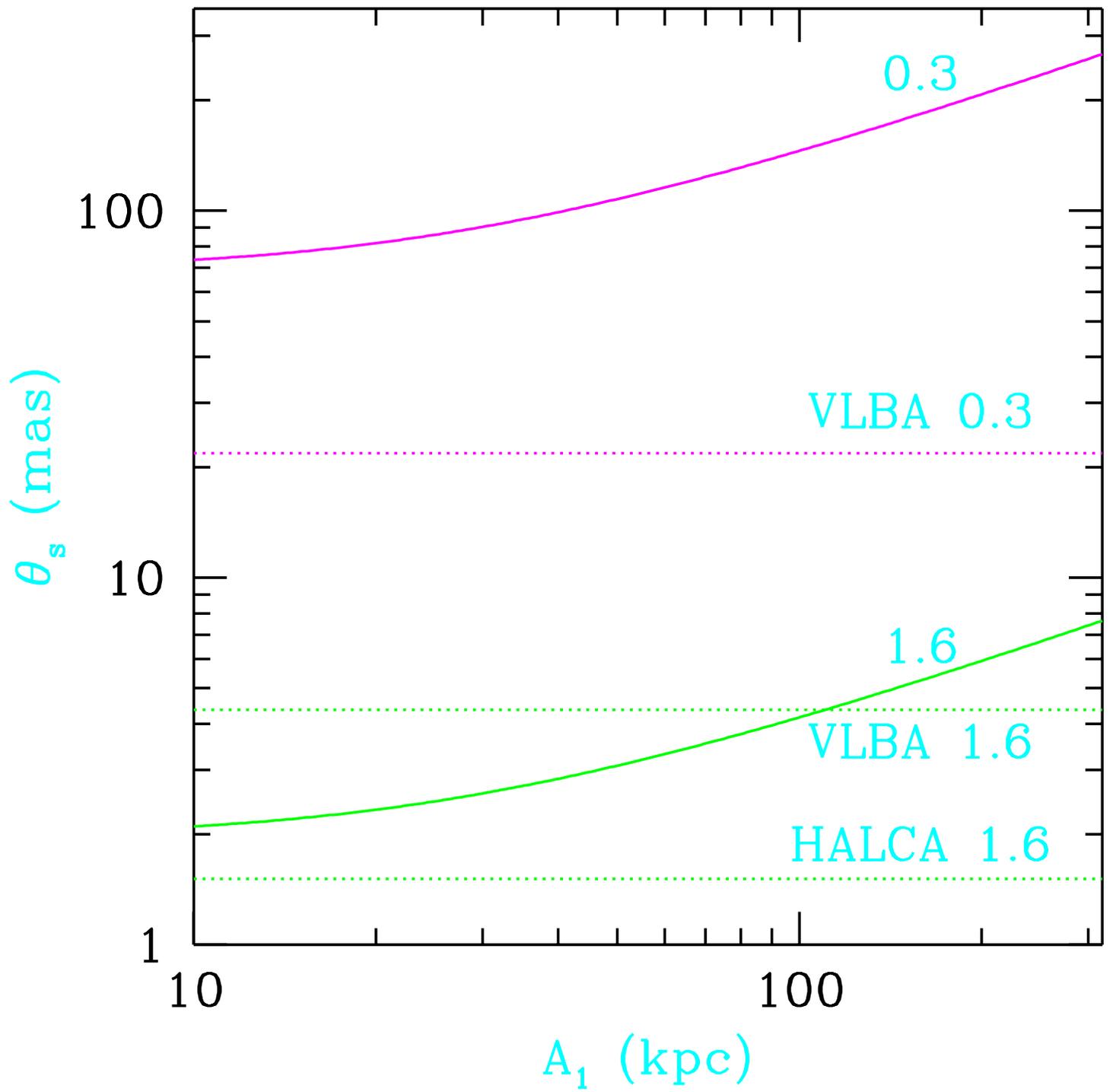
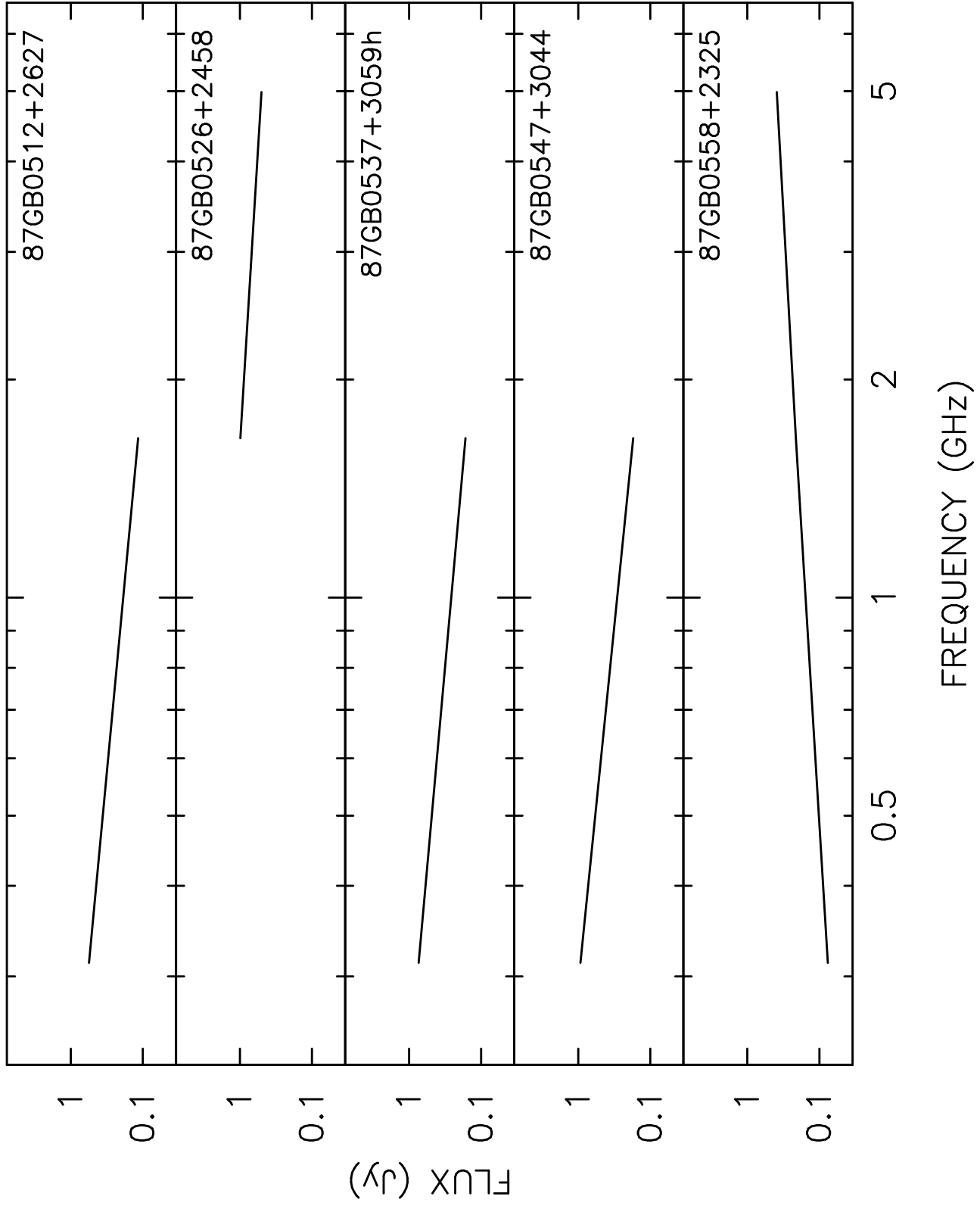
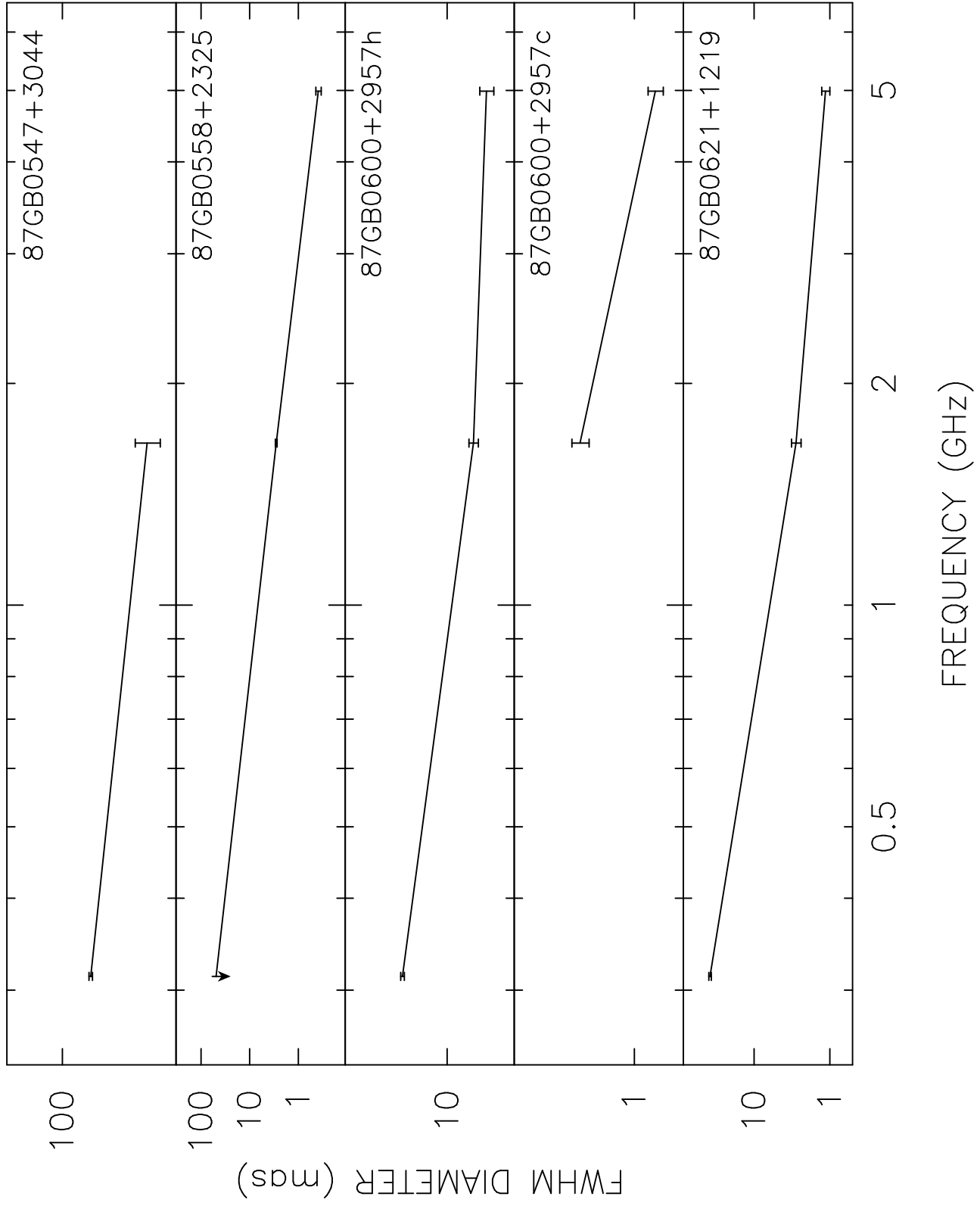
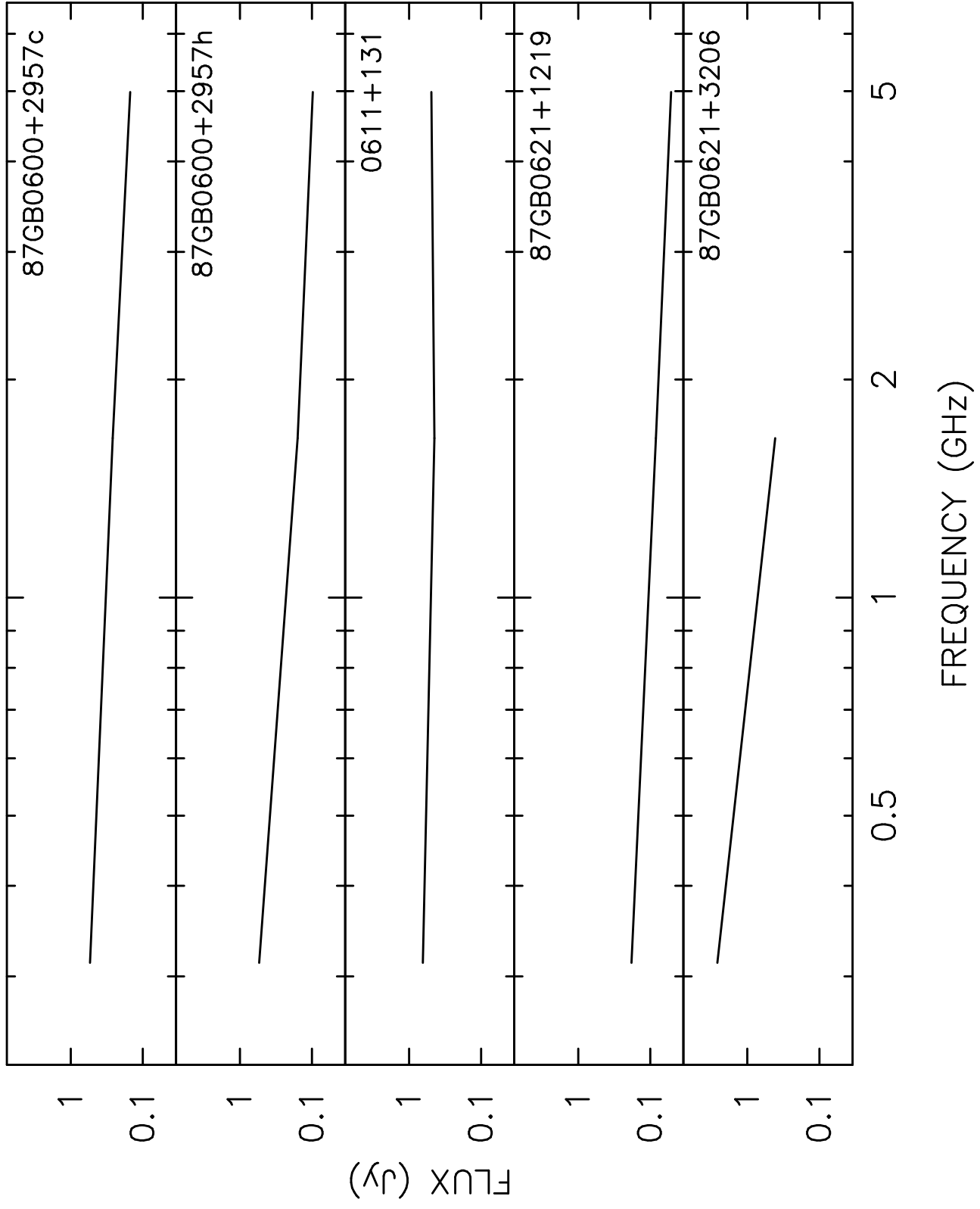
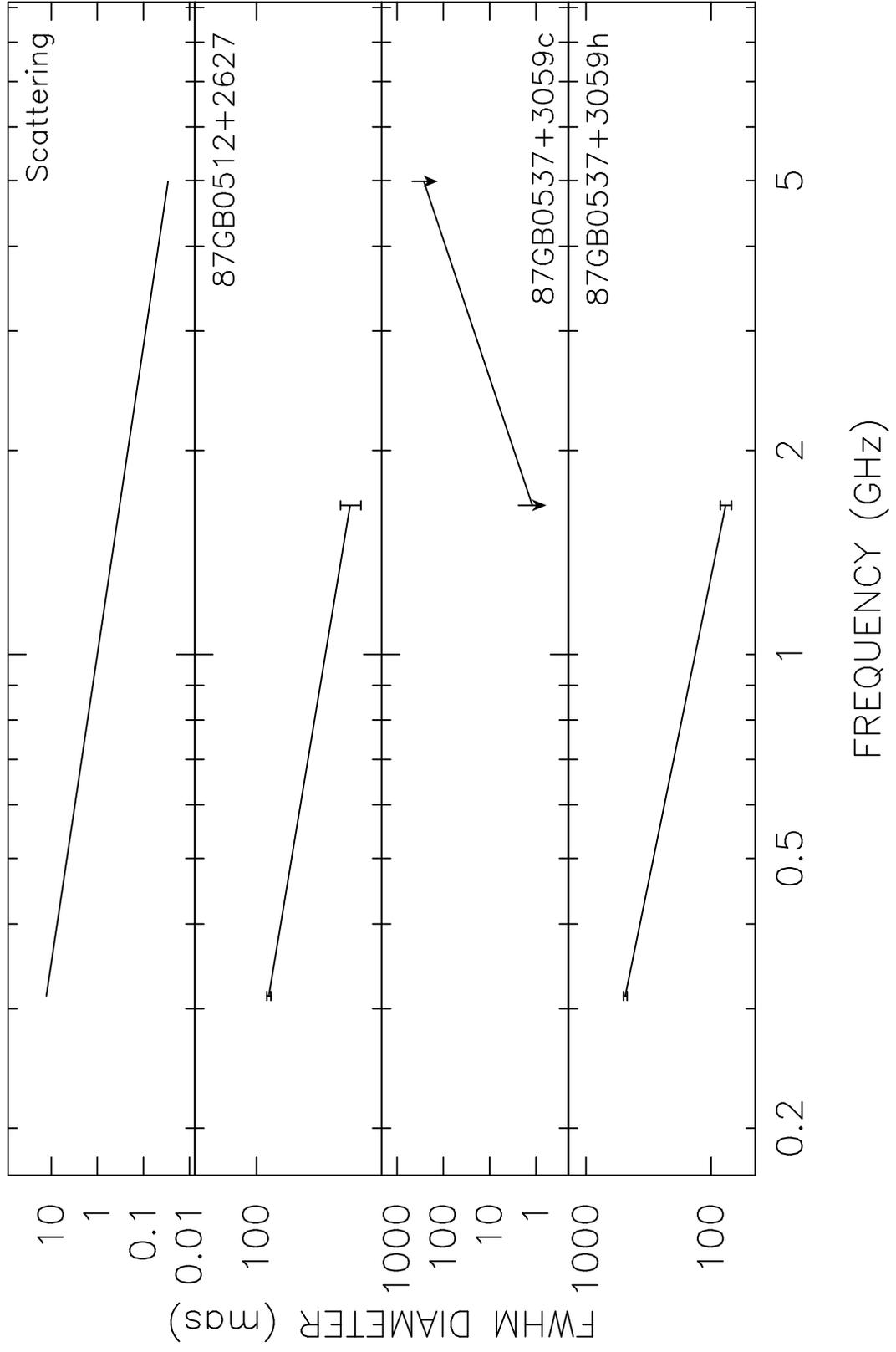
\end{document}